\documentclass[prd, twocolumn, nofootinbib]{revtex4}
\usepackage{epsfig}

\newcommand{\beq}{\begin{equation}}
\newcommand{\eeq}{\end{equation}}
\newcommand{\beqa}{\begin{eqnarray}}
\newcommand{\eeqa}{\end{eqnarray}}
\newcommand{\om}{\Omega_m}

\def\ga{\mathrel{\mathpalette\fun >}}
\def\fun#1#2{\lower3.6pt\vbox{\baselineskip0pt\lineskip.9pt
  \ialign{$\mathsurround=0pt#1\hfil##\hfil$\crcr#2\crcr\sim\crcr}}}

\begin{document} 

\title{On Oscillating Dark Energy} 
\author {Eric V.\ Linder} 
\affiliation{Berkeley Lab, University of California, 
Berkeley, CA 94720} 


\begin{abstract} 
Distance-redshift data can impose strong constraints on dark energy models 
even when the equation of state is oscillatory.  Despite the double 
integral dependence of the distance on the equation of state, precision 
measurement of the distance-redshift relation for $z=0-2$ is more 
incisive than the linear growth factor, 
CMB last scattering surface distance, and the age of the 
universe in distinguishing oscillatory behavior from an average behavior. 
While oscillating models might help solve the coincidence problem 
(since acceleration occurs periodically), next generation observations 
will strongly constrain such possibilities. 
\end{abstract} 


\maketitle 

\section{Introduction \label{sec:intro}} 

The subtle variations in the cosmic expansion hold fundamental 
information on the nature of dark energy and the cosmological model. 
While these variations are not so extreme as to cause nonmonotonicity 
in the scale factor vs.\ time relation $a(t)$, and even less so in its 
integral the distance-redshift relation, we might consider nonmonotonicity 
in the dark energy equation of state $w(a)$.  Having more than one period 
of strongly negative equation of state, and hence acceleration, could 
ameliorate the coincidence problem of why acceleration is happening now 
out of all the expansion history of the universe.  Here we examine the 
implications of periodicity (e.g.\ a single Fourier mode of 
nonmonotonicity) in the equation of state $w(a)$, and what constraints 
can be placed on it. 

First, we note that many periodic or nonmonotonic potentials have been 
put forward for dark energy, but these will not necessarily, and indeed 
rarely, give rise to periodic $w(a)$.  As one well-studied example 
\cite{frieman}, the 
potential for a pseudo-Nambu Goldstone boson (PNGB) field can be written 
as $V(\phi)=V_0[1+\cos(\phi/f)]$, clearly periodic, where $f$ is a 
(axion) symmetry energy scale.   However, unless the field has already 
rolled through the minimum, the relation $w(a)$ is monotonic and indeed 
can be well described by the usual $w(a)=w_0+w_a(1-a)$ (see \S\ref{sec:misc} 
for further discussion).  

For a damped field, we can estimate how far the field has rolled in 
the history of the universe by writing 
\beq 
\dot\phi\sim \sqrt{(1+w)\rho_w}\sim \sqrt{1+w}\ HM_P, 
\eeq 
where $M_P$ is the Planck energy and $H$ is the Hubble parameter. 
If $w$ is close to $-1$, as observations seem to indicate, then the 
field has only traversed 
\beq 
\Delta\phi \sim \dot\phi/H \ll M_P, 
\eeq 
and may well not have had time to see nonmonotonicity in the potential. 
That is, if one walks in darkness only a few paces one may not realize 
there are hills and valleys in the terrain. 

Therefore, rather than examining nonmonotonic potentials and calculating 
periodic behavior in $w(a)$, and then investigating those effects on 
cosmological observables, we start directly with a phenomenological, 
periodic equation of state (see \cite{baren2} for some particle physics 
motivation for such an equation of state and \cite{feng} for its application 
to crossing $w=-1$).  From the calculation of 
effects on observables 
we will impose constraints on the amplitude and period of dark energy 
oscillations. 

\section{Limits on Oscillation} 

Our four basic tools for constraining dark energy equation of state 
oscillations will be the distance-redshift relation out to redshift 
$z=2$ (as can be accurately measured by Type Ia supernovae), the 
distance to the cosmic microwave background last scattering surface 
at $z=1089$, the age of the universe today, and the linear 
growth factor of mass perturbations.  These will constrain the models 
sufficiently that we do not need to extrapolate to and employ early 
time constraints 
from the primordial nucleosynthesis or recombination epochs.  However, 
such early universe constraints are important for oscillatory models 
that are not periodic, such as the tracking oscillating model of 
\cite{dodelson}. 

Note that despite the double 
integration leading from the equation of state to the distance-redshift 
relation, sufficiently precise distance measurements can provide 
strong constraints on periodicity in $w(a)$.  Indeed we find they are 
the most incisive probe. 

Since we take a purely phenomenological model for $w(a)$, we do not have 
information on perturbations in the dark energy.  However, canonical 
scalar fields generically do not have significant perturbations on 
smaller than the Hubble scale, so we will neglect inhomogeneities in 
the dark energy (but see \S\ref{sec:early}).  Note that coupling of 
dark energy to matter 
\citep{amendola,koivisto,groexp,oz} can produce both inhomogeneities and 
nonmonotonic equations of state.  We leave that scenario for future 
investigation, but generally it should cause significant deviations 
in the growth behavior. 

The natural period of the cosmic expansion is given by 
$H^{-1}=(d\ln a/dt)^{-1}$, so we examine periodicity in units of 
the e-folding scale $\ln a$.  That is, our ansatz is 
\beq 
w(a)=w_0-A\sin(B\ln a), \label{eq:w3}
\eeq 
where $w_0$ is the value today. 
We expect that as the central value, $\langle w\rangle\approx w_0$, 
becomes less negative, 
the acceleration decreases and it will be harder to match the distance 
and growth relation of, say, a cosmological constant $w=-1$ universe.  
For example, the undulant universe model \citep{baren3}, with an effective 
$\langle w\rangle=0$, can quickly be ruled out on the basis of 
possessing only 3\% of the growth by today of the cosmological constant 
model \citep{groexp}. 

As the amplitude of oscillations, $A$, increases we likewise expect 
clear distinction from a model with constant $w$.  As the frequency 
of oscillations $B$ increases, however, the observations, which 
depend on integrals of the equation of state, could have difficulty 
distinguishing the oscillatory model from a model with constant equation of 
state $\langle w\rangle$. 

To generally describe a sinusoidal equation of state we would need 
four parameters: 
\beq 
w(a)=w_c-A\sin(B\ln a+\theta), 
\eeq 
where $\theta$ gives the phase of the oscillation today ($a=1$) 
and $w_c$ gives the center of the range over which $w$ oscillates 
(and generally the average value). 
In Eq.~(\ref{eq:w3}), and for the rest of this paper, we assume that 
$\theta=0$ so that $w_c=w_0$.  This should make it more difficult 
to distinguish an oscillating model from one with a constant equation 
of state $w=w_0$, and so represents a conservative approach. 
(Note that the slinky equation of state \cite{baren2} has $\theta=\pi/2$, 
$w_c=0$, $A=1$, while the undulant model further 
specializes to $B=1$.) 

\begin{figure}[!hbt]
\begin{center} 
\psfig{file=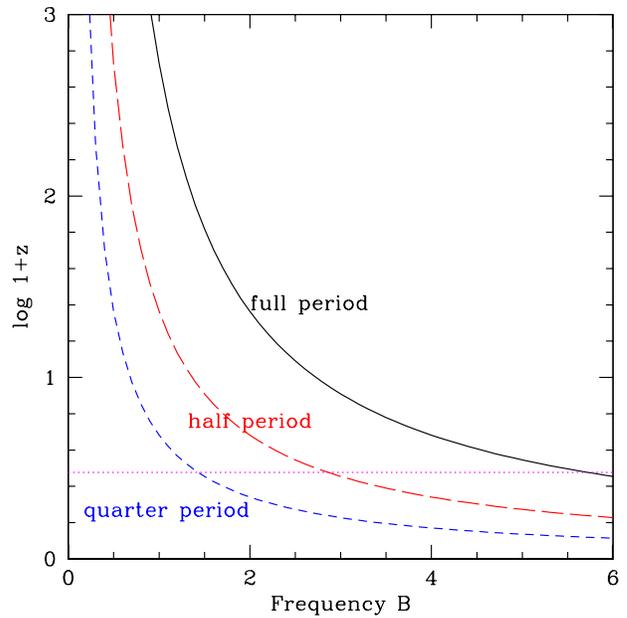,width=3.4in} 
\caption{The frequency $B$ of the oscillatory equation of state 
determines a minimum redshift for seeing a full (or half or quarter) 
period of the oscillations (and hence recognizing a distinction from 
a monotonically varying equation of state).  The magenta, dotted line 
denotes $z=2$, so for example a full oscillation period is only obtained 
from observations within the epoch $z=0-2$ for $B>5.72$. 
}
\label{fig:blimit} 
\end{center} 
\end{figure}

We explore the three remaining parameters one by one to isolate 
the differing aspects of the oscillations.  Of course when the 
amplitude $A=0$ then the frequency $B$ is not defined as we have 
reduced to a constant equation of state.   
To actually see oscillatory behavior, we need $|B\ln a_{\rm min}|>2\pi$, 
where $z_{\rm max}=a_{\rm min}^{-1}-1$ gives the upper redshift bound on 
the precision observations, but we will consider lower frequencies. 
For smaller $B$ 
we might simply see a monotonic trend in $w$ over the region of 
observations.  Figure~\ref{fig:blimit} plots the redshift of the 
first full period of oscillation.  Only for $B>5.72$ does a full period 
occur in the range $z=0-2$.

For a given $w_0$, we can find the maximum amplitude of the oscillation 
consistent with interpreting the observations in terms of a constant, 
averaged equation of state $\langle w\rangle=w_0$.  Thus, larger 
variations could be distinguished as a varying equation of state. 
As observational constraints we employ future data equivalent to 
1\% distance measures over the redshift range $z=0-2$ (e.g.\ from 
SNAP-quality Type Ia supernovae data \cite{snap}), 0.7\% distance measure 
to the CMB last scattering surface at $z=1089$ (e.g.\ from Planck Surveyor 
CMB temperature anisotropy data \cite{planck}), 1\% measure of the 
current age of the universe $H_0t_0$, or 2\% measure of the linear growth 
factor over $z=0-2$ (e.g.\ from SNAP-quality weak lensing data). 
Note that for these probes the cosmological constant ($w=-1$) case lies 
respectively 2.3\%, 1.0\%, 1.8\%, 3.1\% away from the $w=-0.9$ model 
(keeping the matter density $\om=0.28$).  We find that supernovae 
distances provide the most stringent constraints on the oscillation 
amplitude $A$ and frequency $B$. 

\begin{figure}[!hbt]
\begin{center} 
\psfig{file=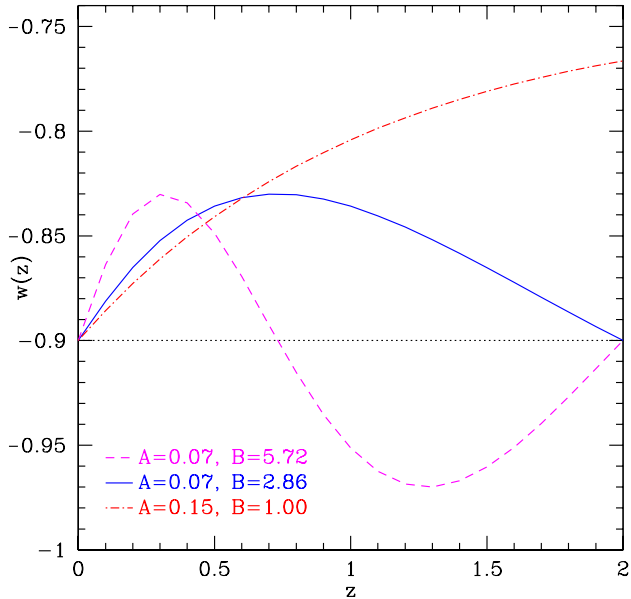,width=3.4in} 
\caption{These equation of state curves possess the maximal 
amplitude $A$ allowed for specific frequencies 
$B$, given next generation observations constraining the magnitude-redshift 
relation to $\Delta m<0.02$ relative to the constant equation of 
state model $w=w_0=-0.9$, for $z=0-2$.   Oscillatory equations of state 
with a given frequency must lie closer to the dotted, constant equation 
of state line than the pictured curves to match the constraints.  
Upon flipping the sign of $A$ (initially decreasing the equation of 
state into the past) these curves are basically reflected about the 
$w=-0.9$ line. 
}
\label{fig:wzosc} 
\end{center} 
\end{figure}

These results are shown in Fig.~\ref{fig:wzosc}, for $w_0=-0.9$ and 
three values of frequency $B$, corresponding to one full period (i.e.\ 
a true oscillation within the range $z=0-2$), one half period, and 
$B=1$.  Larger amplitudes than those shown would violate the observational 
constraints.  The constraints on $A$ are rather symmetric about zero for 
the small amplitudes allowed; that is, the equation of state could 
equally well decrease into the past by the same amount. 

\begin{figure}[!hbt]
\begin{center} 
\psfig{file=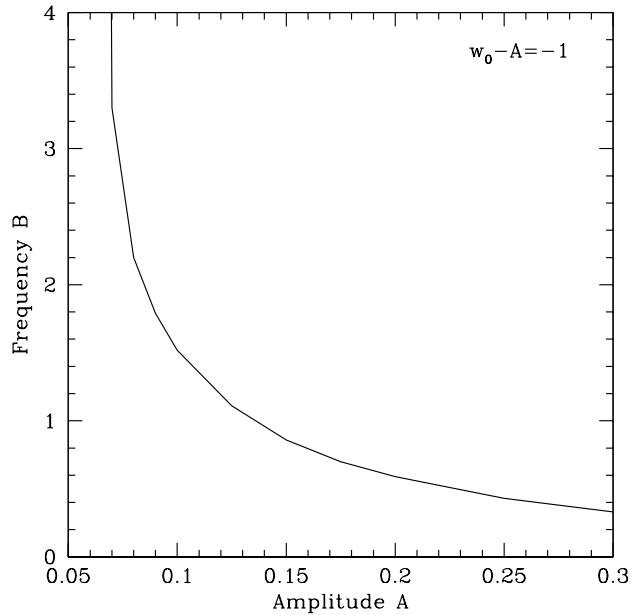,width=3.4in} 
\caption{For a frequency $B$ the curve defines the maximal amplitude 
allowed, given next generation observations constraining the 
magnitude-redshift relation to $\Delta m<0.02$ for $z=0-2$ relative to the 
constant equation of state model $w=w_0=-1+A$.  Such a 
model has an equation of state reaching a minimum value of $-1$, the 
cosmological constant value. 
}
\label{fig:avsb} 
\end{center} 
\end{figure}

As we take $w_0$ less negative, the allowed amplitudes for a given 
frequency decrease.  Basically, since the dark energy density dies off 
less quickly for less negative $w_0$, the oscillations have a 
substantial impact over a larger 
scale factor range, and so constraints are tighter. 
Upon changing $w_0$ we can also adjust the oscillation amplitude $A$ so 
as to keep the same minimum (or maximum) equation of state.  
For example, for a canonical scalar field, 
$w\ge-1$, so if we wish to avoid oscillations violating this bound we 
could require $w_0-A\ge-1$. 
Figure~\ref{fig:avsb} shows the relative constraints on amplitude and 
frequency when the minimum is set to the cosmological constant value $w=-1$. 
For low amplitudes, the oscillations are not discernible in the observations 
so any value of frequency is allowed.  As the amplitude increases, the 
frequency must decrease in order for the oscillatory equation of state 
to remain looking close to a constant equation of state in the observational 
constraints.

\section{Oscillation at Early Times \label{sec:early}} 

Oscillating equations of state can act as models contributing a 
substantial fraction of the total energy density at early times ($z\gg1$). 
These are not really early quintessence models, though, in that the 
equation of state is not necessarily providing acceleration, i.e.\ $w<-1/3$. 
The energy density of the oscillating model is 
\beq 
\rho_{\rm osc}(a)=\rho_{\rm osc}(1)\,a^{-3(1+w_0)} e^{(3A/B)[1-\cos(B\ln 
a)]}. \label{eq:rho} 
\eeq
For the limiting cases of Figs.~\ref{fig:wzosc}-\ref{fig:avsb}, the 
contribution to the 
total energy density at $z=2$ is within 20\% of what a plain $w=-0.9$ 
model would give, and the energy density fraction at $z=1089$ remains 
at the $10^{-8}$ level.  
The slinky model manages 
to have high early energy density -- possibly even greater than the 
matter density -- by taking the central value $w_c=0$.  Thus, early energy 
density is achieved at the price of not maintaining dark energy, i.e.\ 
acceleration. 

Another caution is that early presence of substantial energy density 
not in the matter component can obviate the usual matter dominated epoch. 
While the consequences of this for structure formation are recognized 
in the suppression of the linear growth factor, there is a more subtle 
effect.  The second order differential equation for the growth factor 
$g=(\delta\rho_m/\rho_m)/a$ requires two boundary conditions, canonically 
given by $g(a\ll1)=1$ and $dg/da(a\ll1)=0$ (cf.\ \cite{linjen}).  
However, these arise from 
the usual density growth behavior $\delta\rho_m/\rho_m\sim a$ for a 
matter dominated universe.  In numerical computation of the growth 
equation in an oscillating model, we find a sensitive interplay between 
the initial scale factor $a_i$ and the fraction of energy density at 
that scale factor in components other than matter, such that the 
boundary condition is unstable.  Consequently, for slinky models, 
numerical instability of the growth factor arises for $B\ga0.2$.  This 
may be apparent in the nonmonotonicity of the growth shown in 
Fig.~11 of \cite{baren3}.  Proper solution in this regime requires 
solution of the full set of coupled matter, scalar field, and radiation 
perturbation equations so 
as to give the proper initial conditions (velocity) to the matter growth. 
Note that some initial velocity $dg/da>0$ can actually help offset the 
suppression of the growth factor. 

Models, e.g.\ where $w_c\sim-1$, that do have a true matter dominated 
epoch, do not suffer from these complications. 
Interestingly, if $w_c=w_0\sim -1$, growth data alone allows two 
distinct regions in parameter space.  For low frequencies $B$, the 
equation of state is nearly constant and so the growth mimics that of 
the constant $w$ model.  (If $w_c=-1$, the oscillating model in the 
limit $B\to0$ approaches the cosmological constant.)  
As $B$ increases, the growth is altered due 
to the change in the Hubble parameter relative to the constant $w$ 
model.  The tension between growth and solution of the coincidence 
problem (that acceleration is observed near today) is nicely highlighted by 
\cite{baren3}.  For very high frequencies $B$, the growth once again 
approaches the constant $w$ case (cf.\ Eq.~\ref{eq:rho}).  This holds 
even if the amplitude of oscillations $A\approx1$.  

A high oscillation frequency, though, may be expected to be accompanied 
by spatial inhomogeneities in the dark energy due to the (relatively) 
large effective mass of the field.  This will then alter the matter 
perturbation growth.  Regardless, 
use of precision distance data removes 
the high frequency fit region, for large amplitudes.  For small amplitudes 
any frequency $B$ is allowed by the distance data, as shown in 
Fig.~\ref{fig:wzosc}.

\section{Other Oscillations \label{sec:misc}} 

We briefly return to the issue of oscillations of the potential, or 
the field motion in the potential, rather than oscillations of the 
equation of state directly.  As mentioned in \S\ref{sec:intro}, the 
PNGB model has a periodic potential.  For accelerating behavior, however, 
the field must not have undergone many oscillations, else the equation 
of state would be near zero like the axion case.  

Caldwell \& Linder 
\cite{caldlin} examined the dynamics and found the behavior in the 
$w'-w$ phase plane to be nearly linear: $w'\approx F(1+w)$, where 
$w'=dw/d\ln a$.  The coefficient $F$ is proportional to 
the inverse of the symmetry scale $f$. 
They showed this was a thawing model, where the field starts frozen 
and hence looking like a cosmological constant $w=-1$, before the field 
begins to evolve as it comes to dominate the cosmic energy density. 
The solutions for the equation of state and energy density for such 
dynamics are 
\beqa 
w(a)&=&-1+(1+w_0)a^{F} \\ 
\rho_w(a)&=&\rho_w(1) e^{(3/F)(1+w_0)(1-a^{F})}\ ,
\eeqa 
which are manifestly non-oscillatory.  As mentioned in \S\ref{sec:intro}, 
this looks like the standard varying equation of state $w(a)=w_0+w_a(1-a)$ 
when $F\approx1$, which holds for natural values $f\sim{\cal O}(M_P)$. 

Another possibility is for the potential not to be periodic but the 
field motion is.  Consider oscillations around a minimum of the potential. 
Turner \cite{turner83} showed that for oscillation periods much smaller 
than the Hubble time (large frequency $B$ in our notation), the averaged 
equation of state becomes 
\beq 
\langle w\rangle = \frac{n-2}{n+2}, \label{eq:phin} 
\eeq 
for potentials $V\sim\phi^n$ with $n$ even.  For potentials looking 
like quadratic (quartic) potentials near their minimum, this gives 
$\langle w\rangle=0$ (1/3) -- behavior like matter (radiation).  To 
provide acceleration, one would need $n\ll1$ (conversely, very steep 
potentials give behavior like shear energy, $\langle w\rangle=1$). 
If the frequency of oscillations is comparable or larger than the 
Hubble time (so the energy density changes significantly during 
the oscillations), 
the equation of state does not obey Eq.~(\ref{eq:phin}), but in this 
case we would not discern oscillations in the data. 

Thus, it is not easy to obtain oscillations in the equation of state 
without fine tuning the potential to have particular nonmonotonic 
periodicities, as in the slinky model.  We can analyze the phase plane 
dynamics of our oscillatory model, Eq.~(\ref{eq:w3}), quite simply (also 
see Fig.~9 of \cite{baren3}).  It satisfies the equation 
\beq 
\frac{(w-w_0)^2}{A^2}+\frac{w'^2}{A^2B^2}=1, 
\eeq 
so the dynamics is given by an ellipse in the phase space with center 
at $w=w_0$, $w'=0$, semimajor axis $A$, semiminor axis $AB$, 
and eccentricity $\sqrt{1-B^2}$ (or if $B>1$, switch major and minor,  
and the eccentricity is $\sqrt{1-B^{-2}}$).  
Note that $B=1$, where the characteristic oscillation scale equals the 
Hubble e-fold scale, gives a circle.

\section{Conclusion} 

Precision distance-redshift measurements over the range $z=0-2$ possess 
power in discriminating oscillating equations of state from constant 
ones, despite the double integral relation.  Other probes are not as 
constraining, but growth measurements can provide important limits 
on the early time behavior if the dark energy affects the degree of 
matter domination.  Growth must be calculated carefully, however, since 
high amplitudes affect the growth boundary conditions and numerical 
stability, and high frequencies can lead to spatial inhomogeneities in 
the dark energy.  While oscillating models offer one idea for solving 
the coincidence problem (since acceleration occurs periodically), 
next generation observations will be able to 
strongly constrain such possibilities.

\end{document}